\documentclass[12pt]{article}

\usepackage{float}

\usepackage{graphicx}
\usepackage{epsfig}
\usepackage{amssymb}
\usepackage{color}

\def\gtwid{\mathrel{\raise.3ex\hbox{$>$\kern-.75em\lower1ex\hbox{$\sim$}}}}
\def\ltwid{\mathrel{\raise.3ex\hbox{$<$\kern-.75em\lower1ex\hbox{$\sim$}}}}
\def\square{\kern1pt\vbox{\hrule height 1.2pt\hbox{\vrule width 1.2pt\hskip 3pt
   \vbox{\vskip 6pt}\hskip 3pt\vrule width 0.6pt}\hrule height 0.6pt}\kern1pt}
   
\usepackage{amsmath}

\begin{document}

\begin{titlepage}

\begin{flushright}
UFIFT-QG-19-06, CP3-19-43
%\\
%CP3-19-43
\end{flushright}

\vskip 0.7cm

\begin{center}
{\bf\Large Breaking of scaling symmetry by massless scalar on de Sitter}
\end{center}

\vskip .3cm

\begin{center}
D. Glavan$^{1*}$, S. P. Miao$^{2\star}$, T. Prokopec$^{3\dagger}$ and
R. P. Woodard$^{4\ddagger}$
\end{center}

\vskip .3cm

\begin{center}
\it{$^{1}$ Centre for Cosmology, Particle Physics and Phenomenology (CP3) \\
Universit\'e catholique de Louvain, Chemin du Cyclotron 2, 1348 Louvain-la-Neuve, BELGIUM}
\end{center}

\begin{center}
\it{$^{2}$ Department of Physics, National Cheng Kung University \\
No. 1, University Road, Tainan City 70101, TAIWAN}
\end{center}

\begin{center}
\it{$^{3}$ Institute for Theoretical Physics, Spinoza Institute \& EMME$\Phi$ \\
Utrecht University, Postbus 80.195, 3508 TD Utrecht, THE NETHERLANDS}
\end{center}

\begin{center}
\it{$^{4}$ Department of Physics, University of Florida,\\
Gainesville, FL 32611, UNITED STATES}
\end{center}

\vspace{.5cm}

\begin{center}
ABSTRACT
\end{center}

We study the response of a classical massless minimally coupled scalar to a static point scalar charge 
on de Sitter. By considering explicit solutions of the problem
we conclude that -- even though the dynamics formally admits dilatation 
(scaling) symmetry -- the physical scalar field profile necessarily breaks the symmetry.
This is an instance of symmetry breaking in classical physics due to large infrared effects.
The gravitational backreaction, on the other hand, does respect dilatation symmetry, making this an
example of symmetry non-inheritance phenomenon.

\begin{flushleft}
\end{flushleft}

\vskip .5cm

\begin{flushleft}
$^{*}$ e-mail: drazen.glavan@uclouvain.be \\
$^{\star}$ e-mail: spmiao5@mail.ncku.edu.tw \\
$^{\dagger}$ e-mail: T.Prokopec@uu.nl \\
$^{\ddagger}$ e-mail: woodard@phys.ufl.edu
\end{flushleft}

\end{titlepage}

%\section{Scaling symmetry and massless scalar on de Sitter} 
%\label{Scaling symmetry and massless scalar on de Sitter}

% In this letter we investigate the classical analogue of this phenomenon, namely
% breaking of {\it scaling symmetry} by a classical scalar field $\phi$. 

\noindent{\bf Point particle and scaling solution.}
In this note we investigate the system of a massless minimally coupled scalar (MMCS) field $\Phi$
in de Sitter space coupled to a scalar point charge. The action for the MMCS in an arbitrary 
curved space is given by,
\begin{equation}
S_0[\Phi] =\int \! d^4x \,\sqrt{-g} \, \biggl[
	-\frac{1}{2} g^{\mu\nu}  \! \left(\partial_\mu\Phi\right)\left(\partial_\nu\Phi\right)	\biggr] 
\,, 
\label{scalar field action}
\end{equation}
where~$g^{\mu\nu}$ is the inverse of the metric 
tensor~$g_{\mu\nu}$, $g\!=\!{\rm det}[g_{\mu\nu}]$, the metric signature is $(-,+,+,+)$
and its coupling to the point particle~$\chi^\mu \!=\! \chi^\mu(\tau)$ is modeled by the action,
\begin{equation}
S_{\rm int}[\chi,\Phi] = -
	\int \! d\tau \, \sqrt{-g_{\mu\nu}\dot\chi^\mu(\tau)\dot\chi^\nu(\tau)} \, 
	\lambda\Phi(\chi(\tau)) \, , 
\label{point particle action}
\end{equation}
where~$\tau$ is an affine parameter,~$\lambda$ 
is a dimensionless coupling, and~$\dot\chi^\mu(\tau) \!=\! d\chi^\mu(\tau)/d\tau$. 
We assume that the point particle is at rest, sitting at the origin of the coordinate system
on flat spatial slices of the Poincar\'{e} patch,~$\chi^\mu(\tau) \!=\! (\tau,0,0,0)$. The equation of
motion for the MMCS descends from variation of the action~(\ref{scalar field action}) and~(\ref{point particle action}),
\begin{equation}
 \square \Phi(x)
 	= -\frac{1}{a^2} \Bigl( \partial_0^2 + 2 a H \partial_0 
 		- {\nabla}^2 \Bigr) \Phi(x)
 		=  \lambda \frac{\delta^3(\vec x)}{a^3} 
\, ,
\label{covariant equation of motion}
\end{equation}
where~$a(\eta)\!=\! -1 / (H\eta)$ is the scale factor of de Sitter space with $\eta$ confrormal time,
$H=(\partial_0a)/a^2$ the (constant) Hubble rate, $\partial_0=\partial/\partial \eta$ and $ {\nabla}^2$
is the Laplacian.
While the sourceless equation would respect all of the isometries of de Sitter,
the point source~(\ref{covariant equation of motion}) breaks spatial special conformal transformations 
and spatial translations, leaving us with only {\it four} isometries, namely spatial rotations 
and dilatations, $x^\mu \!\to\! e^{\alpha} x^\mu$ with $\alpha\in\mathbb{R}$.

It is most natural to assume that the solution of~(\ref{covariant equation of motion}) satisfies the background isometries,
and that it depends only on the rotation-invariant and dilatation-invariant combination of
coordinates~$X\!=\!aHr$, $r=\|\vec x\|$, {\it i.e.} $\Phi(\eta,\vec x) \rightarrow \Phi(X)$, 
also known as the {\it scaling solution},
upon which the equation of motion~(\ref{covariant equation of motion}) away from the origin turns into an
ordinary one,
\begin{equation}
\Biggl[ (1 \!-\! X^2) \frac{d}{d X}
	+ \frac{2}{X} (1 \!-\! 2X^2)
	 \Biggr] \frac{d}{d X} \Phi(X) = 0 \, .
\label{massless scaling eq}
\end{equation}
This equation is integrated straightforwardly, and its general solution is
\begin{equation}
\Phi(X) = -\frac{\lambda H}{4\pi X} - \frac{\lambda H}{8\pi} 
	\ln\biggl( \frac{1\!-\!X}{1\!+\!X} \biggr)
	+ \Phi_0 \, .
\label{scaling solution}
\end{equation}
One integration constant is completely fixed by the $\delta$-function source term by means of 
the Green's integral theorem, while the remaining trivial constant~$\Phi_0$ remains undetermined.

A closer examination of the solution in~(\ref{scaling solution}) reveals some worrisome
features. Most notably, the solution exhibits a logarithmic 
singularity at the horizon! At a first glance there seems to be nothing wrong with our 
assumptions. Perhaps it is that strong infrared effects that are known to exist for
MMCS in de Sitter conspire to create, in a manner of speaking, a classical wall of fire -- a barrier 
at which the geodesic equation for a test particle becomes singular.

That~(\ref{scaling solution}) cannot be a physical solution can be seen by considering the energy-momentum tensor, 
$T_{\mu\nu}=\partial_\mu\Phi \partial_\nu\Phi
-\frac12 g_{\mu\nu}g^{\alpha\beta}\partial_\alpha\Phi \partial_\beta\Phi$, 
accompanying the solution~(\ref{scaling solution}), which in spherical coordinates reads,
\begin{equation}
{T^\mu}_\nu = H^2 
\begin{pmatrix}
-\frac{1}{2}(1\!+\! X^2)\!	&	 -  X 	&	0	&	0
\\
 X 	&	\!\frac{1}{2}(1\!+\! X^2)\!	&	0	&	0
\\
0	&	0	&	\!-\frac{1}{2}(1\!-\!X^2)\!	&	0
\\
0	&	0	&	0	&	\!-\frac{1}{2}(1\!-\!X^2)
\end{pmatrix}\!
\Bigl( \frac{\partial \Phi}{\partial X} \Bigr)^2
\,.
\label{energy momentum}
\end{equation}
Near the horizon it diverges quadratically, as can be easily seen from,   
\begin{equation}
\Bigl(\frac{\partial \Phi}{\partial X}\Bigr)^2
    \;\;\overset{X\to1}{\sim}\;\;\frac{\lambda^2 H^2}{64\pi^2}\frac{1}{(1\!-\!X)^2}
\,.
\end{equation}
This divergence of the diagonal terms would generate
a large classical backreaction onto the background space-time.
In particular, there is a positive radial energy density flux~${T^r}_0$, 
which also diverges quadratically at the horizon.
While the divergence at the origin $\propto 1/(ar)^4$ is the usual divergence 
generated by a point charge 
that is dealt with in the usual way, the divergence at the Hubble horizon cannot be a 
part of the physical solution.
In order to shed light on the origin of the problem, in the next section we consider the 
equivalent problem 
for a massive scalar and construct a solution that is regular everywhere except at the origin.

\bigskip

\noindent{\bf Massive scalar on de Sitter.} 
A massive scalar field satisfies the equation of motion,
\begin{equation}
\Bigl( \square - m^2 \Bigr) \Phi(x) 
	= \lambda\frac{\delta^3(\vec{x})}{a^3} \, .
\label{eom: massive}
\end{equation}
This equation still possesses dilatation symmetry, and thus admits a scaling solution
that away from the origin satisfies a homogeneous equation,
\begin{equation}
\Biggl[ (1 \!-\! X^2) \frac{d^2}{d X^2}
	+ \frac{2}{X} (1 \!-\! 2X^2) \frac{d}{d X}
	- \frac{m^2}{H^2}
	 \Biggr] \Phi(X) = 0 \, .
\end{equation}
The general solution can be written in terms of two hypergeometric functions,
\begin{eqnarray}
&& \hspace{-0.7cm}
\Phi(X)  =
	-\frac{\lambda H}{4\pi X} \times
	{}_2F_1\biggl( \biggl\{ \frac{1}{4} \!+\! \frac{\nu}{2} , \frac{1}{4} \!-\! \frac{\nu}{2} \biggr\} , 
		\biggl\{ \frac{1}{2} \biggr\} , X^2 \biggr)
\nonumber \\
&&	\hspace{0.7cm}
	+ \frac{\lambda H}{2\pi}  \times
	\frac{\Gamma\bigl( \frac{3}{4} \!+\! \frac{\nu}{2} \bigr) \, 
				\Gamma\bigl( \frac{3}{4} \!-\! \frac{\nu}{2} \bigr)}
		{\Gamma\bigl( \frac{1}{4} \!+\! \frac{\nu}{2} \bigr) \,
				\Gamma\bigl( \frac{1}{4} \!-\! \frac{\nu}{2} \bigr)} \times
	{}_2F_1\biggl( \biggl\{ \frac{3}{4} \!+\! \frac{\nu}{2} , \frac{3}{4} \!-\! \frac{\nu}{2} \biggr\} , 
		\biggl\{ \frac{3}{2} \biggr\} , X^2 \biggr) \, ,
\qquad 
\label{massive solution}
\end{eqnarray}
where,
\begin{equation}
\nu = \sqrt{ \frac{9}{4} \! - \! \frac{m^2}{H^2} } \, .
\label{nu}
\end{equation}
The constant in front of the first hypergeometric function is fixed by the source in~(\ref{eom: massive}), 
while the second one is fixed by the requirement of regularity at the horizon.
Moreover, the behaviour of the solution for~$X\!\to\!\infty$ is regular.
One can add  to~(\ref{massive solution}) a homogeneous solution that breaks scaling symmetry,
but such contributions tend to be subdominant at late times.

Examining the result~(\ref{massive solution}) in the small mass limit is instructive for understanding
the issues involved in the massless scaling solution~(\ref{scaling solution}),
\begin{eqnarray}
&&
\hspace{-0.5cm}
\Phi(X)  \overset{m\to0}{\sim}
	-\frac{\lambda H}{4\pi X} \,
	- \frac{\lambda H}{8\pi} \ln \biggl( \frac{1 \!-\! X}{1 \!+\! X} \biggr)
\nonumber \\
&&	\hspace{1cm}
	- \frac{\lambda H}{2\pi}  \biggl[ \frac{3H^2}{2m^2} + \ln(2) - \frac{7}{6} \biggr]
	+ \frac{\lambda H}{8\pi} \ln\bigl(1 \!-\! X^2 \bigr)
%	+{\mathcal{O}}(m^2)
		 \, , 
\qquad
\label{massless limit}
\end{eqnarray}
The first line in this expansion comes from the first line of the full 
solution~(\ref{massive solution}), and reproduces the massless 
solution~(\ref{scaling solution}) up to a constant. The second line above 
comes from the small mass expansion of the second line in~(\ref{massive solution}).
It is clear there is no singularity at the horizon even in this limit.
However, it is also clear that this limit is singular
due to the constant term~$\sim1/m^2$. One might try to employ the observation
that the massless solution~(\ref{scaling solution}) is defined up to
a constant in order to remove the divergent term above. This though does
not work, as~(\ref{massless limit}) with the divergent constant removed simply
does not satisfy the massless equation of motion~(\ref{massless scaling eq}).
The proper conclusion is that the scaling
solution of our problem is singular in the massless limit, and~(\ref{scaling solution})
does not represent a valid physical solution. In other words, there is no scaling solution 
for the massless case that is regular away from the origin.

The physical interpretation of this behavior
is clear: the point source generates a large amount of classical infrared scalar modes such
that it breaks scaling (dilatation) symmetry in the limit of small mass. This is
the reason behind why the na\^ive scaling solution~(\ref{scaling solution}) we found in the massless
case has a pathological behavior at the horizon.  The small mass behavior 
in~(\ref{massless limit}) is reminiscent of the well understood massless limit
of the MMCS propagator in de Sitter space, which we briefly recap in the following.

\bigskip

\noindent{\bf Scalar propagator in de Sitter.}
The small mass behavior in~(\ref{massless limit}) 
is reminiscent of the better known example in linear quantum physics in
de Sitter space. There exists a de Sitter invariant two-point Wightman function 
for a massive scalar in 
de Sitter~\cite{Chernikov:1968zm},
\begin{equation}
\bigl\langle \hat{\phi}(x) \hat{\phi}(x') \bigr\rangle
	\!=\! \frac{H^2\, \Gamma\bigl( \frac{3}{2} \!+\! \nu \bigr) \,
	\Gamma\bigl( \frac{3}{2} \!-\! \nu \bigr)}{(4\pi)^2} 
	\times {}_2F_1 \biggl(
	\biggl\{ \frac{3}{2} + \nu , \frac{3}{2} - \nu \biggr\} , \biggl\{ 2 \biggr\} , 
	1 - \frac{y}{4}
	\biggr) \, ,
\label{massive Wighman function on de Sitter}
\end{equation}
where~$\nu$ is again the one from~(\ref{nu}), and~$y$
is the de Sitter invariant function of the coordinates, 
%which is for the positive frequency Wightman function~(\ref{massive Wighman function on de Sitter}),
%
\begin{equation}
y(x;x') = a(\eta) a(\eta') H^2 \Bigl[ \| \vec{x} \!-\! \vec{x}^{\,\prime} \|^2 
	- \bigl( \eta\!-\! \eta' \!-\! i\varepsilon \bigr)^2 \Bigr] \, .
\end{equation}
The small mass expansion of this expression is,
\begin{equation}
\bigl\langle \hat{\phi}(x) \hat{\phi}(x') \bigr\rangle
	\overset{m\to0}{\sim}
	\frac{H^2}{(2\pi)^2 } \biggl[ \,
	\frac{1}{y}
	- \frac{1}{2} \ln(y)
	+  \frac{3 H^2}{2 m^2}
	+ \ln(2)-\frac{11}{12}  \, \biggr] \, ,
\end{equation}
which tells us there is no physical and finite de Sitter invariant solution for the massless scalar field
due to strong infrared effects. However, demanding that the state respects only
spatial homogeneity and isotropy yields a perfectly physical 
behavior~\cite{Linde:1982uu,Starobinsky:1982ee,Vilenkin:1982wt,Ford:1984hs,Allen:1985ux,Allen:1987tz,Onemli:2002hr,Onemli:2004mb},
\begin{equation}
\bigl\langle \hat{\phi}(x) \hat{\phi}(x') \bigr\rangle
	=
	\frac{H^2}{(2\pi)^2 } \biggl[ \,
	\frac{1}{y}
	- \frac{1}{2} \ln(y)
	+  \frac{1}{2} \ln(aa')
	+1-\gamma_E \biggr] \, ,
\label{massless scalar propagator}
\end{equation}
where the (non-universal) constant  is fixed by taking the $D=4$ limit of the massless scalar propagator 
from~\cite{Glavan:2015ura}.
This lesson prompts us to look for a physical solution in the case at hand which does not respect
the background isometries to resolve the conundrum.

\bigskip

\noindent{\bf Breaking of dilatation symmetry.} 
\label{Breaking of scaling symmetry}
Here we derive the solution of~(\ref{covariant equation of motion}) by using the Green's function method.
Let us assume that the scalar point charge starts
acting on the scalar field at some initial moment of time~$\eta_0$. We use the method
of Green's function to determine the reaction of the scalar field to this charge.
The retarded Green's functions for a massless scalar field on de Sitter space 
can be straightforwardly obtained from~(\ref{massless scalar propagator}), 
\begin{equation}
G_R(x;x') = 
	- \frac{\theta(\Delta\eta)}{2\pi}
	\biggl[
	\frac{\delta\bigl( \Delta\eta^2 \!-\! \| \Delta\vec{x} \|^2 \bigr)}{a(\eta) \, a(\eta')}
	+
	\frac{H^2}{2} \theta\bigl( \Delta\eta\!-\! \| \Delta\vec{x} \| \bigr)
	\biggr] \, ,
\end{equation}
where~$\Delta\eta\!=\! \eta\!-\! \eta'$, and~$\Delta\vec{x}\!=\! \vec{x}\!-\! \vec{x}^{\,\prime}$.
The scalar potential that solves~(\ref{covariant equation of motion})
is now obtained by integrating the retarded Green's function against
the point source,
\begin{equation}
\Phi(\eta,r) = \int_{\eta_0}^{0}\! d\eta' \int \! d^3x' \, a^4(\eta') \,
	G_{R}(x;x') \, \times \lambda\frac{\delta^3(\vec{x}')}{a^3(\eta')} \, ,
\end{equation}
which evaluates to,
\begin{equation}
\Phi(\eta,r) = \theta\bigl( \eta\!-\!\eta_0 \!-\!r \bigr) \biggl[ -\frac{\lambda H}{4\pi X}
	- \frac{\lambda H}{4\pi} \ln \biggl( \frac{a}{1 \!+\! X} \biggr) \biggr] 
	\, ,
\label{Green's function solution}
\end{equation}
where~$\eta_0=-1/H$ such that $a(\eta_0)=1$. 
The step function in front of the solution accounts
for causality, restricting the effect of the interaction to within the forward light cone of the source. 
Of course, Green's second identity includes surface integrations of the Green's function (and its derivative) times the solution (and its derivative) on the initial value surface. Eq.~(\ref{Green's function solution}) 
has implicitly assumed that the solution and its first time derivative vanish at $\eta = \eta_0$. 
It is more natural to take the initial values from the term inside the square brackets, 
in which case the solution becomes,
\begin{equation}
\Phi(\eta,r) = -\frac{\lambda H}{4\pi X}
	- \frac{\lambda H}{4\pi} \, \ln \biggl(\frac{a}{1 \!+\! X} \biggr) %+\Phi_0
\, .
\label{Green's function solution:2}
\end{equation}
From the point of view of a local observer on de Sitter, Eq.~(\ref{Green's function solution:2})
 is valid on the entire manifold.
The solution~(\ref{Green's function solution:2}) is the {\it principal result} of this letter.
It can be obtained by adding to~(\ref{scaling solution}) a homogeneous solution, 
$\Phi_h=\frac{\lambda H}{8\pi}\ln[(1\!-\!X^2)/a^2]$, 
resulting in a solution that is {\it regular everywhere} except at the origin. 
However, the scaling symmetry is broken by the term $\propto \ln(a)$. 
 It should be noted that at late times
and at large radial separations the dominant contribution is time-independent and grows
logarithmically with the comoving distance,
\begin{equation}
\Phi(\eta,r) \overset{r\to\infty}{\sim}
	\frac{\lambda H}{4\pi} \ln\bigl( Hr \bigr) \, .
\end{equation}
The energy-momentum tensor for~(\ref{Green's function solution:2}) reads,
\begin{equation}
{T^\mu}_\nu \!=\! \frac{\lambda^2H^4}{32\pi^2}\!
\begin{pmatrix}
\!- \Theta^2\!-\!\Psi^2\!	&	 -2\Theta\Psi 	&	0	&	0
\\
2\Theta\Psi  	&	\!\Theta^2\!+\!\Psi^2\!	&	0	&	0
\\
0	&	0	&	\!\!\!\Theta^2\!-\!\Psi^2\!\!	&	0
\\
0	&	0	&	0	&	\!\!\Theta^2\!-\!\Psi^2\!
\end{pmatrix}
\!,\;\; 
\label{energy momentum:2}
\end{equation}
with $\Psi\!=\!\frac{1}{X^2}\!+\!\frac{1}{1+X}, \Theta\!=\!X\Psi\!-\!1$.
It is regular everywhere away from the origin
and decays as $\sim \!1/X^2$ for large radial distances. Remarkably, 
the energy-momentum tensor in~(\ref{energy momentum:2})
respects dilatation symmetry, even though the field profile in~(\ref{Green's function solution:2}) does not.
This is a cosmological example 
of the phenomenon of (perturbative) {\it symmetry non-inheritance}, which has attracted
significant attention in recent literature~\cite{Herdeiro:2014goa,Smolic:2015txa,Cvitan:2015aha,Smolic:2016dmh,Barjasic:2017oka}.

\bigskip

\noindent{\bf Summary and discussion.} 
\label{Summary and discussion}
We investigate the classical response of a massless scalar field to a static point-like scalar charge on de Sitter.
The point charge breaks spatial special conformal isometries, as well as spatial translations of de Sitter space. 
The resulting equation~(\ref{covariant equation of motion}) possesses only four isometries, namely
spatial rotations and dilatations, also known as global scaling transformations. 
We show that any solution that respects all four isometries
exhibits a logarithmic singularity at the Hubble horizon,
making this na\^ive solution~(\ref{scaling solution})  unphysical.
Inspired by the quantum case of a massless scalar propagator on de Sitter, 
we then show that the classical physical solution~(\ref{Green's function solution:2}) necessarily breaks scaling symmetry and it is regular everywhere except at the point charge location. 
Remarkably, the energy-momentum tensor associated with this solution {\it does} respect dilatation symmetry. 
Therefore, our solution provides an example of the phenomenon of symmetry 
non-inheritance in gravitational systems~\cite{Herdeiro:2014goa,Smolic:2015txa,Cvitan:2015aha,Smolic:2016dmh,Barjasic:2017oka}.
Our analysis can be generalized to 
$D$ space-time dimensions, in which case the na\^ive scaling solution also exhibits a logarithmic singularity at the 
horizon,~\footnote{The scaling solution which generalizes~(\ref{scaling solution}) to $D$ dimensions is,
\begin{equation}
\Phi(X)=-\frac{\lambda H^{D-3}\Gamma\left(\frac{D-3}{2}\right)}{4\pi^\frac{D-1}{2}X^{D-3}}
  \times_{2}\!F_1\biggl( \biggl\{ 1 , \frac{3\!-\!D}{2}  \biggr\} , 
		\biggl\{  \frac{5\!-\!D}{2}  \biggr\} , X^2 \biggr)
+\Phi_0\,,
\nonumber
\end{equation}
which exhibits a logarithmic singularity at the horizon, 
\begin{equation}
\Phi(X) \;\overset{X\to 1}{\sim}\;
	-\frac{\lambda H^{D-3}\,\Gamma\big(\frac{D-1}{2}\big)}{4\pi^{\frac{D-1}{2}}} \ln\bigl(1\!-\!X\bigr)
	\,. 
\nonumber
\end{equation}
}
and therefore the physical solution must break scaling symmetry in arbitrary number of dimensions.

It would be of interest to study physical consequences of such a classical breaking of scaling symmetry,
and in particular whether there are observable late time effects of this symmetry breaking.
For example, our solution can be helpful for improving our understanding 
of how point charges in inflation affect temperature fluctuations in the cosmic microwave background 
radiation~\cite{Carroll:2008br,Prokopec:2010nm}.

 After the completion of this work it came to our attention that some of our results, including equation~(\ref{Green's function solution:2}), 
but not the breakdown of the dilatation invariant 
solution~(\ref{scaling solution}), were previously obtained by 
Akhmedov, Roura and Sadofyev~\cite{Akhmedov:2010ah}.

\bigskip

\noindent{\bf Acknowledgements.}
We are grateful to Ivica Smoli\'c for a critical reading of the manuscript
and for drawing our attention to the symmetry non-inheritance phenomena.
This work was partially supported by the Fonds de la Recherche 
Scientifique -- FNRS under Grant IISN 4.4517.08 -- Theory of fundamental
interactions, by Taiwan MOST grants 103-2112-M-006-001-MY3 
and 107-2119-M-006-014; by the D-ITP consortium, a program of 
the Netherlands Organization for Scientific Research (NWO) that is 
funded by the Dutch Ministry of Education, Culture and Science 
(OCW); by NSF grants PHY-1806218 and PHY-1912484; 
and by the Institute for Fundamental Theory at the University of Florida.

\end{document}